# A Discussion on the Heisenberg Uncertainty Principle from the Perspective of Special Relativity

Luca Nanni

*E-mail:* luca.nanni@student.unife.com

**Abstract** In this note, we consider the implications of the Heisenberg uncertainty principle (HUP) when computing uncertainties that affect the main dynamical quantities, from the perspective of special relativity. Using the well-known formula for propagating statistical errors, we prove that the uncertainty relations between the moduli of conjugate observables are not relativistically invariant. The new relationships show that, in experiments involving relativistic particles, limitations of the precision of a quantity obtained by indirect calculations may affect the final result.

**Abstract** Dans cette note, nous considérons les implications du principe d'incertitude de Heisenberg (HUP) lors du calcul des incertitudes qui influent sur les principales grandeurs dynamiques, du point de vue de la relativité restreinte. En utilisant la formule bien connue pour la propagation des erreurs statistiques, nous prouvons que les relations d'incertitude entre les modules des observables conjugués ne sont pas invariant relativistes. Les nouvelles relations montrent que, dans les expériences impliquant des particules relativistes, les limites de la précision d'une quantité obtenue par des calculs indirects peuvent affecter le résultat final.

# 1   Introduction

The Heisenberg uncertainty principle (HUP) is one of the pillars of quantum theory. Direct consequences of this principle are the commutation relations that limit the simultaneous measurement of the conjugate observables of motion, and the inability to measure a quantity with arbitrary accuracy without perturbing the physical reality of the system [1,2,3,4]. Moreover, the HUP can explain quantum behavior phenomena like tunneling (which manifests in natural phenomena such as nuclear radioactivity or fusion in main sequence stars) or the creation of matter-antimatter couples [5,6]. In fact, the experimental uncertainties of some conjugate variables such as energy and time may generate significantly wide fluctuations in the quantum states to produce (in the probabilistically sense) phenomena that are not allowed in classical mechanics. The HUP means that the world of the very small is inherently elusive. In this sense, the uncertainty principle assumes an epistemological value because it limits what we can know about a quantum particle.

However, the literature does not contain a formal reworking of the HUP for special relativity, except for the paper of Rosen from 1932 [7]. In this article, the author used a purely physical approach to prove that the uncertainty relations between conjugate variables are also valid from the perspective of special relativity. Research has recently been published on the minimal position-velocity uncertainty based on the dispersion of a wave packet associated to a relativistic free particle [8]. This paper proved that the minimal uncertainty relation is frame-dependent.

The main representations of the HUP commonly encountered in scientific literature are relative to the couples of position-momentum, energy-time, and action-angle. That is,

$$\begin{cases} \mathrm{dqdp} \geq \hbar/2, \\ \mathrm{dEdt} \geq \hbar/2, \\ \mathrm{dSd\theta} \geq \hbar/2, \end{cases}$$

where q and p are the components of the position and linear momentum vectors, respectively.

A didactic discussion on the implications of special relativity on the HUP was presented by Landau et al. in the 4th volume of the course of Theoretical Physics [9]. However, their analysis was restricted to calculating the lower limit of the uncertainty of the measured observable. That is,

$$\begin{cases} \mathrm{dqdp} = (v' - v)\mathrm{dtdp} \geq \frac{\hbar}{2} & \Rightarrow & \mathrm{dp} \geq \frac{\hbar}{2\mathrm{dt}(v'-v)} \\ \mathrm{dqdp} = \mathrm{dqdE}/c \geq \frac{\hbar}{2} & \Rightarrow & \mathrm{dq} \geq \frac{\hbar c}{2\mathrm{dE}} \end{cases}, \quad (1)$$

where v and v′ are the velocities of the system (particle) before and after the measurement. These relations imply that, after fixing the error of one of the two conjugate variables, there is a lower limit on the uncertainty that is always non-zero, i.e.,

$$\begin{cases} (\mathrm{dp})_{\min} \geq \lim_{(v'-v) \to c} \frac{\hbar}{2\mathrm{dt}(v'-v)} = \frac{\hbar}{2c\mathrm{dt}} \\ (\mathrm{dq})_{\min} \geq \lim_{\mathrm{dE} \to mc^2} \frac{c\hbar}{2\mathrm{dE}} = \frac{\hbar}{2mc} = \frac{\hbar}{2p} \end{cases}. \quad (2)$$

The second line of **(2)** represents the ultra-relativistic limit, where the minimum error of the position is half the De Broglie wavelength [9]. Therefore, from the relativity perspective, the uncertainty of a measurement of a conjugate variable can never be reduced to zero, even if it is precise. In particular, the uncertainty of the position can never be lower than the order of the De Broglie wavelength, whereas the uncertainty of the linear momentum can only be reduced to zero by performing an infinitely long measurement (which has no physical meaning). These limitations on the uncertainties of conjugate variables are also propagated to the dynamical quantities that they are directly or indirectly used to calculate. Moreover, if the uncertainty principle is extended to the moduli of the vectors position and momentum, then the relativistic effects change the structure of the relations in **(2)**. In the following, we formulate the uncertainty relationship for the moduli of conjugate observables, and discuss how they affect the precision of the calculated dynamical quantities.

## 2  The HUP for the Modulus of Conjugate Variables: Non-Relativistic Case

Consider a free quantum particle and suppose that we wish to measure all the components of its position and linear momentum vectors. The measurement is a classical process and we wish to measure all of the components at the same time. We let $dp_i$ and $dq_i$ denote the statistical errors that affect the vectorial components. The results of the measurement are

$$\mathbf{q} = (q_1 \pm dq_1, q_2 \pm dq_2, q_3 \pm dq_3) = (q_1, q_2, q_3) \pm (dq_1, dq_2, dq_3),$$

and

$$\mathbf{p} = (p_1 \pm dp_1, p_2 \pm dp_2, p_3 \pm dp_3) = (p_1, p_2, p_3) \pm (dp_1, dp_2, dp_3).$$

The absolute value of the dot product of the vectors formed by the uncertainties of the single position and momentum components (considering the Heisenberg uncertainty relations) is

$$|\langle d\mathbf{p}, d\mathbf{q}\rangle| = |\langle (dp_1, dp_2, dp_3), (dq_1, dq_2, dq_3)\rangle| =$$

$$= \left|\sum_{i=1}^{3} dq_i dp_i\right| \geq \left|\frac{3}{2}\hbar\right| > \frac{\hbar}{2}. \tag{3}$$

Recall the Holder inequality [10] given by

$$|\langle \mathbf{v}, \mathbf{w}\rangle|_{p+q} \leq \|\mathbf{v}\|_p \|\mathbf{w}\|_q \qquad \frac{1}{p} + \frac{1}{q} = 1, \tag{4}$$

which represents the generalization of the Cauchy-Schwarz inequality. Here, **v, w** are vectorial functions belonging to a given space $L^p(\Omega)$ (space of measurable functions that are p-th power integrable). In our case, p and q are equal to 2 ($L^2(\Omega)$ is then a Hilbert space). Inequality **(4)** implies that

$$|\langle d\mathbf{p}, d\mathbf{q}\rangle| \leq \|d\mathbf{p}\|\|d\mathbf{q}\| = \sqrt{\sum_{i=1}^{3} dq_i^2 dp_i^2}, \qquad (5)$$

where the sums under square roots are performed using Einstein formalism. Because we have proved that $|\langle \delta\mathbf{p}, \delta\mathbf{q}\rangle| > \hbar/2$, the results in **(3)** and **(5)** imply that

$$\|d\mathbf{p}\|\|d\mathbf{q}\| = dpdq \geq \frac{\hbar}{2}. \qquad (6)$$

Therefore, the Heisenberg inequality that is typically applied to the vectorial components of conjugate observables may be extended to the modulus of the vectors representing these variables in the Euclidean space (that is, the space where the measurements occur). We can get the same result by applying the uncertainty propagation formula. In this respect, we calculate the moduli of vectors **p** and **q** using

$$\|\mathbf{p}\| = \sqrt{\sum_{i=1}^{3} p_i} \qquad \text{and} \qquad \|\mathbf{q}\| = \sqrt{\sum_{i=1}^{3} q_i}.$$

The uncertainties affecting these two quantities are obtained by differentiation, that is,

$$dp = \sqrt{\sum_{i=1}^{3}\left(\frac{\partial \|\mathbf{p}\|}{\partial p_i} dp_i\right)^2} = \frac{1}{\|\mathbf{p}\|}\sqrt{p_1^2 dp_1^2 + p_2^2 dp_2^2 + p_3^2 dp_3^2}$$

and

$$dq = \sqrt{\sum_{i=1}^{3}\left(\frac{\partial \|\mathbf{q}\|}{\partial q_i} dq_i\right)^2} = \frac{1}{\|\mathbf{q}\|}\sqrt{q_1^2 dq_1^2 + q_2^2 dq_2^2 + q_3^2 dq_3^2}.$$

Multiplying the uncertainties and recalling the Heisenberg uncertainty relationships we get

$$dpdq = \frac{1}{\|\mathbf{p}\|}\frac{1}{\|\mathbf{q}\|}\sqrt{p_1^2 q_1^2 dq_1^2 dp_1^2 + p_2^2 q_2^2 dq_2^2 dp_2^2 + p_3^2 q_3^2 dq_3^2 dp_3^2} =$$

$$\geq \frac{\hbar}{2}\frac{1}{\|\mathbf{p}\|}\frac{1}{\|\mathbf{q}\|}\sqrt{\|\mathbf{p}\|^2\|\mathbf{q}\|^2} = \frac{\hbar}{2},$$

which proves the previous statement. Equation **(6)** is used in the next section.

## 3   The HUP for the Modulus of Conjugate Variables: Relativistic Case

Consider a free quantum particle with rest mass $m_0$ that is moving with relativistic velocity v. Assume that all the position and momentum components have been measured and are known with their uncertainties. We let x denote the modulus of the relativistic vectors position, and let p denote the linear momentum of the particle, i.e.,

$$x = c\tau = \frac{ct}{\gamma} = \sqrt{c^2 t^2 - q^2} \tag{7}$$

and

$$p = \gamma m_0 v \tag{8}$$

where $\gamma = (1 - v^2/c^2)^{-1/2}$. Moreover, we let δ be the uncertainty of the relativistic variable and d be the uncertainty of its space component. Applying the formula for propagating the statistical errors to **(7)** and **(9)** we get

$$(\delta x)^2 = \left(\frac{ct}{\sqrt{c^2 t^2 - q^2}}\right)^2 dt^2 + \left(\frac{q}{\sqrt{c^2 t^2 - q^2}}\right)^2 dq^2$$

and

$$(\delta p)^2 = (\gamma m_0)^2 dv^2 + \left(\frac{m_0 v^2}{\sqrt{(1-v^2/c^2)^3}}\right)^2 dv^2 = \left(\frac{\gamma c^2}{c^2 - v^2}\right)^2 dv^2.$$

Multiplying these two square uncertainties and recalling that $dE = cdp$ and $q = ct\sqrt{1 - 1/\gamma^2}$, we can use simple calculations to get

$$(\delta x)^2 (\delta p)^2 = \gamma^8 (dE)^2 (dt)^2 + \gamma^8 \left(\sqrt{1 - 1/\gamma^2}\right)^2 (dq)^2 (dp)^2. \tag{9}$$

The product of the square uncertainties of conjugate relativistic observables is given by the linear combination of their square time and space modulus components. The coefficients of the linear combinations are not constant but are frame-dependent. Recalling **(6),** we can rewrite **(9)** as

$$(\delta x)^2 (\delta p)^2 \geq \gamma^8 \left(\frac{\hbar}{2}\right)^2 + \gamma^8 \left(\sqrt{1-\frac{1}{\gamma^2}}\right)^2 \left(\frac{\hbar}{2}\right)^2,$$

from which we can obtain the final version of the uncertainty relation for the relativistic modulus of the position and linear momentum:

$$\delta x \delta p \geq \frac{\hbar}{2} \gamma^4 \sqrt{\left(2-\frac{1}{\gamma^2}\right)}. \tag{10}$$

In a low speed regime (v ≪ c), the relativistic coefficient (γ) tends to one and **(10)** becomes

$$\lim_{\gamma \to 1} \frac{\hbar}{2} \gamma^4 \sqrt{\left(2-\frac{1}{\gamma^2}\right)} = \frac{\hbar}{2}.$$

So, as expected, at non-relativistic velocities **(10)** reduces to the non-relativistic relation in **(6)**. To the limit of relativistic speeds, where γ tends to infinity, **(10)** can be simplified to

$$v \cong c \quad \Rightarrow \quad \delta x \delta p \geq \frac{\sqrt{2}}{2} \hbar \gamma^4. \tag{11}$$

We conclude that from the special relativity perspective, the product of the uncertainties (δrδp) progressively increases with the speed.

The relativistic effect only becomes significant when the speed of the quantum particle is very close to that of light. For instance, when the speed is 98.5 % of the speed of light, $\zeta = \gamma^4 \sqrt{\left(2-\frac{1}{\gamma^2}\right)}$ is in the order of $10^3$. Increasing the speed to more than 99 % of the speed of light increases ζ so that it is in the order of $10^6$. All the relativistic quantities calculated using the moduli of the position and momentum vectors are affected by statistical errors, which are influenced by **(11)**.

# 4 How HUP and the Relativistic Effects May Affect Quantum Tunneling

In the previous section, we proved that the relativistic relation **(11)** between the statistical errors that affect the moduli of position and momentum vectors is only applicable to quantum phenomena that occur in very extreme conditions (high energy physics). This means that most of the main measurements typically performed in atomic or nuclear fields may be analyzed using the HUP or **(6)**. However, we studied how the relativistic effects can affect some typical quantum phenomena such as tunneling.

We now rework the product of relativistic uncertainties, recalling that

$$\tau = \gamma t$$

and

$$\epsilon = \gamma m_0 \Delta v^2 = p \Delta v.$$

We suppose that $\Delta v$ is very close to the speed of light. Thus,

$$\delta r \delta p = \delta(\tau \Delta v)\delta\left(\frac{\varepsilon}{\Delta v}\right) = \delta\varepsilon\delta\tau = \delta\varepsilon\gamma dt.$$

Substituting this result into **(6),** we get

$$\delta\varepsilon dt \geq \frac{\hbar}{2}\gamma^3\sqrt{\left(2 - \frac{1}{\gamma^2}\right)}, \quad \textbf{(12)}$$

or using the simplified form in **(11), (12)** becomes

$$\delta\varepsilon dt \geq \frac{\sqrt{2}}{2}\gamma^3 \hbar. \quad \textbf{(13)}$$

The relativistic uncertainty of the energy of states is

$$\delta\varepsilon \geq \frac{\sqrt{2}}{2dt}\gamma^3\hbar. \qquad (14)$$

This is a generalization of the Landau inequality [8], which can be easily obtained by the second part of **(1)**. That is,

$$dq = \Delta v\, dt \geq \frac{\hbar c}{2dE} \quad \Rightarrow \quad dE \geq \frac{\hbar c}{2dt\Delta v}.$$

In the limit of $\Delta v \cong c$, the inequality becomes

$$dE \geq \frac{\hbar}{2dt}. \qquad (15)$$

The difference between **(14)** and **(15)** is given by the relativistic coefficient $\sqrt{2}\gamma^3$, which increases the energy fluctuation of the states. Inequality **(15)** states that, for quantum systems that approach the limit of the speed of light, high fluctuations of the energy of states can only be obtained by performing measurements over a very short time. According to **(14),** the same result can be obtained or further enhanced by increasing the particle speed. Therefore, phenomena such as quantum tunneling may be obtained by the measuring time and the energy of the interaction with the measuring instrument (for example, the energy of the photon used as a projectile), and are particularly expected in the relativistic limit.